\documentclass[twocolumn,showpacs,amsmath,amssymb]{revtex4}

\usepackage{graphicx}
\usepackage{dcolumn}
\usepackage{bm}

\begin{document}

\title{Muon catalysis of superheavy element production in
nucleus-nucleus fusion reaction}

\author{V.Yu. Denisov}
\email{v.denisov@gsi.de; denisov@kinr.kiev.ua}
 \affiliation{%
Gesellschaft f\"ur Schwerionenforschung (GSI), Planckstrasse 1,
D-64291 Darmstadt, Germany \\ Institute for Nuclear Research,
Prospect Nauki 47, 03028 Kiev, Ukraine}%

\date{\today}

\begin{abstract}
It is shown that muon bound with light projectile induces the
superheavy elements production in nucleus-nucleus collisions.
\end{abstract}

\pacs{25.60.Pj, 
25.70.Jj, 
25.85.-w, 
36.10.-k 
}
\maketitle

\section{Introduction}

The synthesis of superheavy elements (SHEs) was and still is an
outstanding research object. The production cross section of SHEs
with $Z \geq 112$ is very low and close to the limit of current
experimental possibility [1-4]. Due to this it is of interest to
find new types of reactions, which can induce fusion of two heavy
nucleus. We show in the next section that muon bound with light
nucleus induce SHE formation during nucleus-nucleus fusion
reactions.

It is easy to understand qualitatively a influence of muon $\mu^-$
on the SHE fusion process, if we recollect that the wave function
of $1s$ state of $\mu^-$ in a very heavy nucleus is located inside
the nucleus \cite{kim}. Therefore, negatively-charged muon inside
heavy nucleus should effectively reduce the Coulomb repulsion
between protons. Due to this the forces, inducing fission of
compound nucleus and preventing fusion of two nuclei should
decrease. Consequently, the SHE formation probability should rise
due to $\mu^-$.

\section{Catalysis of the SHE synthesis by muon}

The process of SHE formation is subdivided into three steps
\cite{dh,d_she}. 1. The capture of two nuclei in an
entrance-channel potential well and formation of a common nuclear
system of two touching nuclei. 2. The formation of a spherical or
nearly spherical compound nucleus during shape evolution from the
common nuclear system of two touching nuclei to a compound
nucleus. 3. The surviving of the excited compound nucleus due to
evaporation of neutrons and $\gamma$-ray emission in competition
with fission. The capture process depends on both the barrier
thickness and pocket shape of entrance-channel potential between
nuclei \cite{dn}. The shape evolution step is determined by
potential-energy landscape between the touching configuration of
two colliding nuclei and the compound nucleus \cite{dh,d_she}.
Decay properties of the compound nucleus drastically depend on the
fission barrier height \cite{dh,d_she}. Therefore, enhancement of
the SHE production in fusion reaction may be achieved by processes
which (1) make capture pocket dipper and barrier of
entrance-channel potential thinner, (2) increase the slope or
reduce both barrier height and thickness of the potential-energy
landscape between touching configuration of two colliding nuclei
and compound nucleus, (3) increase the fission barrier height.
Below we show that these three conditions can be met in a reaction
between a light nucleus with captured $\mu^-$-meson $L_\mu$ and a
heavy nucleus $T$.

The potential energy of $L_\mu$+$T$ system before touching can be
approximated as
\begin{equation}
E_{L_\mu T}(R) = B_L + B_T + B_{L\mu} + V_{LT}(R) + V_{T\mu}(R),
\end{equation}
where $B_L$ and $B_T$ are the binding energies of light $L$ and
heavy $T$ nuclei, respectively, $B_{L\mu}$ is the binding energy
of muon in the light nucleus $L$, $V_{LT}(R)$ is the interaction
potential between the light and heavy nuclei related to Coulomb
and nuclear forces at distance $R$ between their mass centers
\cite{dn,d_pot}, and $V_{T\mu}(R)=-e^2 Z_T/R$ is the Coulomb
interaction between $Z_T$ protons in the heavy nucleus and the
muon.

The potential energy of the compound nucleus with bound $\mu^-$ is
connected with the binding energy of the compound nucleus $B_{CN}$
and with that of muon in the compound nucleus $B_{CN\mu}$, i.e.,
\begin{equation}
E_{CN} = B_{CN} + B_{CN\mu} .
\end{equation}

The potential energy evaluated relatively to the ground state of
compound nucleus with bound $\mu^-$, which formed during
$L_\mu$+$T$ fusion reaction, is related to difference
\begin{equation}
\delta(R)=E_{L_\mu T}(R)-E_{CN}.
\end{equation}
It is useful to split $\delta(R)$ into contributions of pure
nuclear $\delta_N(R)$ and muon-nuclear $\delta_{N\mu}(R)$
subsystems
\begin{equation}
\delta(R) = \delta_N(R) + \delta_{N\mu}(R) ,
\end{equation}
where
\begin{eqnarray}
\delta_N(R) = B_L + B_T -  B_{CN} + V_{LT}(R), \\
\delta_{N\mu}(R) = B_{L\mu} - B_{CN\mu} + V_{T\mu}(R).
\end{eqnarray}
We see in (3)-(6) that the Coulomb interaction between muon and
protons modifies the potential-energy landscape of  fusing system.
(Note that realistic landscape of potential-energy surface of
fusing system depends on a great number of various collective
coordinates. However, in (1),(3)-(6) we take into account only the
most important collective coordinate, which describes the distance
between mass centers of separated nuclei or elongation of fusing
system upon the capture step.) At distance $R_{CN}$, which
corresponds to the mass distance between left and right parts of
compound nucleus, $\delta_N(R_{CN})=\delta_{N\mu}(R_{CN})=0$. If
$\delta_{N\mu}(R)$ continuously decreases with reducing of $R$,
then muon induces the SHE formation due to three effects. (1) A
more dipper capture pocket is formed as a result of such $R$
dependence of $\delta_{N\mu}(R)$. Therefore, the capture state
formation probability increases. (2) The potential-energy
landscape of the muon-nuclear system becomes more favorable for
shape evolution from captured states of two touching nuclei to the
compound nucleus. (3) The muon-nuclear system exhibits a larger
fission barrier height as compared to pure nuclear system, see
also \cite{mu_fis} and papers cited therein. Consequently, the
fission or quasi-fission probability of muon-nuclear system get
reducing as compared to the pure nuclear system.

In Table 1 we evaluate $\delta_{N\mu}(R)$ around barrier for
several colliding systems, which can be used to the SHE
production. The muon binding energies in muonic atoms is obtained
by using Pustovalov parametrization \cite{kim,mu_b_e} with nucleus
charge radius $R=1.2 A^{1/3}$ fm, where $A$ is the number of
nucleons in the nucleus. The entrance-channel fusion barriers
$B_{gs}$ and $B_{gs \; \mu}$, depths of entrance-channel capture
potential well $D_{pw}$ and $D_{pw \; \mu}$ and bottom of capture
pocket $B_{cp}$ and $B_{cp \; \mu}$ for the pure nuclear $B_{gs}$
and muon-nuclear systems, respectively, are given in Table 1.
Quantities $B_{gs}$, $B_{gs \; \mu}$, $B_{cp}$ and $B_{cp \; \mu}$
are evaluated relatively to the compound-nucleus ground state. The
parameters of entrance-channel potential wells for pure nuclear
systems $B_{gs}$, $D_{pw}$ and $B_{cp}$, presented in Table 1, are
taken from \cite{dn}.

\begin{table}[htb]
\caption{Parameters of the entrance-channel potential for nuclear
and muon-nuclear systems. Quantities $B_{gs}$, $B_{gs \; \mu}$,
$D_{pw}$, $D_{pw \; \mu}$, $B_{cp}$, $B_{cp \; \mu}$ and
$\delta_{N\mu}(R)$ are given in MeV, $R$ in $\delta_{N\mu}(R)$ is
given in fm.}
\begin{tabular}{lcccc}
\hline \hline
Pure nuclear reaction & $B_{gs}$ & $D_{pw}$ & $B_{cp}$ &  \\
Muon-nuclear reaction & $B_{gs \; \mu}$ & $D_{pw \; \mu}$
& $B_{cp \; \mu}$ & $\delta_{N\mu}(R)$\\
\hline  \vspace{-3mm}\\
$^{70}$Zn$+^{208}$Pb$\rightarrow$ $^{278}112$ & 16.7 & 4.0 & 12.7 &  \\
$^{70}$Zn$_\mu$+$^{208}$Pb$\rightarrow$ $^{278}112_\mu$ & 20.3 &
5.3 & 15.0 & $13.19 - \frac{118.08}{R}$ \\ \vspace{-3mm}\\
\hline  \vspace{-3mm}\\
$^{78}$Ge$+^{208}$Pb$\rightarrow$ $^{286}114$ & 12.8 & 2.9 & 9.9 & \\
$^{78}$Ge$_\mu$+$^{208}$Pb$\rightarrow$ $^{286}114_\mu$ & 16.4 &
3.8 & 12.6 & $13.21 - \frac{118.08}{R}$ \\ \vspace{-3mm}\\
\hline  \vspace{-3mm}\\
$^{86}$Kr$+^{208}$Pb$\rightarrow$ $^{294}118$ & 5.3 & 1.1 & 4.2 & \\
$^{86}$Kr$_\mu$+$^{208}$Pb$\rightarrow$ $^{294}118_\mu$ & 9.1 &
1.9 & 7.2 & $13.28 - \frac{118.08}{R}$ \\ \vspace{-3mm}\\
\hline \hline
\end{tabular}
\end{table}

The muon induce fusion reactions, because as we see in Table 1
$\mu^-$ provides (1) an the increase of the entrance-channel
pocket depth $D_{pw \; \mu} > D_{pw}$, (2) a change of the
potential energy landscape from the touching configuration of two
colliding nuclei to the spherical compound nucleus due to $B_{gs
\; \mu} > B_{gs}$ and $B_{cp \; \mu} > B_{cp}$ and (3) an increase
of the fission barrier height. Reasons for increasing both the
fission barrier height and $B_{gs \; \mu}$ are similar, but for
deformed nuclear shape, like in the case of fission barrier shape,
the amplitude of fission barrier increase is smaller, then that
for fusion barrier height. Therefore, all sequential stages of SHE
formation during fusion reaction are induced by muon.

Relative influence of muon on the capture well and
potential-energy landscape between the touching configuration of
two colliding nuclei to the compound nucleus shape is stronger for
the case of system with shallow pocket, as for reaction
$^{86}$Kr$_\mu+^{208}$Pb$\rightarrow$ $^{294}118$, see in Table 1.
Therefore muon catalysis may give possibility to use successfully
reactions with shallow capture pocket.

The height of fusion barrier for muon-nuclear system, evaluated
relatively the ground state of colliding system $L_\mu+T$, is
lower then the one for pure nuclear system  $L+T$ on $\approx -Z_T
e^2/R_b$, where $R_b$ is the distance between nuclei at fusion
barrier. It is necessary to take into account this shift at the
choice of collision energy for $L_\mu+T$ system.

\section{Conclusions}

$\mu^-$ is a convenient particle for inducing compound-nucleus
formation in reactions $L_\mu + T \rightarrow SHE + xn + e^- +
\overline{\nu}_e + \nu_\mu$, because its lifetime ($\approx
2.2\times 10^{-6}$s \cite{pp}) is sufficient for making $1s$ bound
state with a light projectile nucleus just before the collision
with a target and induce fusion reaction. The process of SHE
formation during nucleus-nucleus collision is fast relatively
typical $\mu^-$ dynamic time. Therefore there is high probability
of population of $1s$ bound state of $\mu^-$ in SHE during nuclear
reaction time. Due to this we can use estimates for $\delta_{N
\mu}$, presented in Table 1. The compound nucleus relatively
rarely excited during the decay $\mu^-$ ($\mu^- \rightarrow e^- +
\overline{\nu}_e + \nu_\mu$ \cite{pp}). It is possible to make
beam of muonic projectile $L_\mu$ by merging beams of strongly
ionized projectile nucleus $L$ and of $\mu^-$ at the same
velocities before the target. At such conditions nuclei should
capture $\mu^-$ with high probability and $\mu^-$ should quickly
populates $1s$ state.

Note that muon catalysis of thermonuclear reactions is also
related to effective reduction of the Coulomb repulsion between
protons and is well studied both theoretically and experimentally
(see \cite{kim} and papers cited therein). Muon catalysis of
thermonuclear reactions between two hydrogen isotopes is mainly
related to reduction of both fusion barrier heights and thickness.
In contrast to this muon catalysis of SHE production is connected
with more complex processes as reduction of fusion barrier
thickness, modification of capture pocket, variation of
potential-energy landscape between capture and compound-nucleus
shapes and rising of fission barrier height. The reduction of
fusion barrier height, evaluated relatively to the ground state of
two colliding nuclei, is also taken place in SHE production
reactions with muonic projectiles.

Here we have briefly discuss main features of the muon catalysis
of SHE production. Detail theoretical and experimental studies of
the muon catalysis of heavy nucleus formation are needed.

\section*{Acknowledgements}
The author would like to thank W. N\"orenberg and K.A. Bugaev for
useful discussions and Yu.B. Ivanov for help. Author gratefully
acknowledges support from GSI.

\end{document}